\documentclass[aps,showpacs,preprintnumbers,prd,twocolumn,nofootinbib]{revtex4}

\IfFileExists{srcltx.sty}{\usepackage[active]{srcltx}}

\usepackage{epsfig}
\usepackage{graphicx}
\usepackage{epsfig}
\usepackage{bm}
\usepackage{amsmath}
\usepackage{amssymb}

\newcommand{\beq}{\begin{equation}}
\newcommand{\eeq}{\end{equation}}
\newcommand{\bea}{\begin{eqnarray}}
\newcommand{\eea}{\end{eqnarray}}

\def\keV{\: {\rm keV}}
\def\MeV{\: {\rm MeV}}
\def\GeV{\: {\rm GeV}}

\newcommand{\slashed}[1]{{#1}\hspace{-2mm}/}

\def\simle{\lower 2pt \hbox {$\buildrel < \over {\scriptstyle \sim }$}}
\def\simge{\lower 2pt \hbox {$\buildrel > \over {\scriptstyle \sim }$}}

\begin{document}

\preprint{UCLA/07/TEP/27}

\title{Dark-matter sterile neutrinos in models with a gauge singlet in
the Higgs sector}

\author{Kalliopi Petraki and Alexander Kusenko
}

\affiliation{Department of Physics and Astronomy, University of California, Los
Angeles, CA 90095-1547, USA }


\begin{abstract}

Sterile neutrino with mass of several keV can be the cosmological dark
matter, can explain the observed velocities of pulsars, and can play an
important role in the formation of the first stars.   We describe the
production of sterile neutrinos in a model with an extended Higgs sector, in
which the Majorana mass term is generated by the vacuum expectation value of a
gauge-singlet Higgs boson.  In this model the relic abundance of sterile
neutrinos does not necessarily depend on their mixing angles, the
free-streaming length can be much smaller than in the case of warm dark matter
produced by neutrino oscillations, and, therefore, some of the previously
quoted bounds do not apply.  The presence of the gauge singlet in the Higgs
sector has important implications for the electroweak phase transition,
baryogenesis, and the upcoming experiments at the Large Hadron Collider and a
Linear Collider.

\end{abstract}


\maketitle

\section{Introduction}

The discovery of the neutrino masses can  easily be incorporated into the
Standard Model (SM) by adding two or more SU(3)$\times$SU(2)$\times$U(1)
singlet fermions, often called right-handed neutrinos, which are allowed to
have the Yukawa couplings to the Higgs boson and the standard,  left-handed
neutrinos.  The Yukawa couplings generate the Dirac mass terms for the
neutrinos after the spontaneous symmetry breaking.   In addition, the singlet
fermions can have some Majorana masses. The interplay between the Dirac mass
and the Majorana mass, known as the seesaw mechanism~\cite{seesaw}, can
accommodate the observed neutrino masses for a variety of Majorana masses.   If
the Majorana mass terms are large, the particles associated with the singlet
fields are very heavy. However, if one or more Majorana masses are below the
electroweak scale, the so called sterile neutrinos appear among the low-energy
degrees of freedom.  These new particles can be the cosmological dark
matter~\cite{dw,production_oscillations,shi_fuller,nuMSM,shaposhnikov_tkachev,
Kusenko:2006rh,Kadota:2007mv}, their production in a supernova can
explain the pulsar kicks~\cite{pulsars} and can affect the supernova explosion
in a variety of ways~\cite{supernova_misc}; the same particles can play an
important role in the formation of the first stars~\cite{reion} and some other
astrophysical phenomena~\cite{Biermann:2007ap}.

The properties of the sterile dark matter, and, in particular, how warm or cold
it is for a given mass, depend on the production mechanism.  One mechanism,
which generates a population of relic sterile neutrinos at the sub-GeV
temperature was proposed by Dodelson and Widrow (DW)~\cite{dw}.  If the lepton
asymmetry is negligible, this scenario appears to be in conflict with a
combination of the X-ray bounds~\cite{x-rays} and the
Lyman-$\alpha$ bounds~\cite{viel,silk}.  This conclusion is based on the
state-of-the-art calculations of the sterile neutrino production in neutrino
oscillations~\cite{production_oscillations}.  It is possible to evade this
constraint if the lepton asymmetry of the universe is greater than
$O (10^{-3})$~\cite{shi_fuller}.  On the other hand, some
astronomical observations~\cite{dSphs,cdm-wdm} point to a
non-negligible free-streaming length for dark matter, which favors warm dark
matter.  Moreover, warm dark matter can cause filamentary structure on small
scales~\cite{Gao:2007yk}, in contrast with cold dark matter.  It is also
possible that the sterile neutrinos make up only a fraction
of dark matter~\cite{silk}, in which case they can still be responsible for the
observed velocities of pulsars~\cite{pulsars,Kusenko:2006rh}.

Dodelson--Widrow mechanism is not the only mechanism by which sterile dark
matter could be produced.  The relic population of sterile neutrinos could
be generated in a variety of ways, for example, from their coupling to
the inflaton~\cite{shaposhnikov_tkachev}, the electroweak-singlet Higgs
boson~\cite{Kusenko:2006rh}, or the radion~\cite{Kadota:2007mv}. Whatever the
production history of sterile neutrinos might be at the high temperature, there
is always some additional amount produced in neutrino
oscillations at some sub-GeV temperatures~\cite{dw,production_oscillations}.
The two components can have very different momentum distributions. Therefore,
generically this form of dark matter is a mixed two-component dark matter,
which can have some very interesting observable
consequences~\cite{boyan_mixed}.

In this paper we concentrate on the possibility that the relatively
light Majorana mass could arise via the Higgs mechanism in a model with an
SU(2)$\times$U(1)-singlet Higgs boson coupled to the Standard Model Higgs
boson~\cite{Chikashige:1980ht}, and that the sterile neutrinos could be
produced from the Higgs decays at a temperature as high as
100~GeV~\cite{Kusenko:2006rh}.  We will explore various scenarios for such
production and the implications for the electroweak phase transition.  In
particular, we will address the cooling and the red-shifting of dark matter,
which have important implications for dark matter profiles in
halos~\cite{cdm-wdm,dSphs}, the small-scale structure inferred from
Lyman-$\alpha$ observations~\cite{viel}, and the velocity dispersion in dwarf
spheroids~\cite{dSphs}.

\section{Majorana masses from an extended Higgs sector}
Although the Standard Model was originally formulated with
massless neutrinos $\nu_i$ transforming as components of the
electroweak SU(2) doublets $L_\alpha$ ($\alpha =1,2,3$), the
neutrino masses can be accommodated by a relatively minor
modification.  One adds several electroweak
singlets $ N_{a}$ ($a=1,...,n$) to the Standard Model and builds a
seesaw lagrangian~\cite{seesaw}:
\beq  {\cal L} = {\cal L_{\rm SM}} + i \bar
N_a \slashed{\partial} N_a - y_{\alpha a} H^{\dag} \,  \bar L_\alpha
N_a - \frac{M_a}{2} \; \bar N_a^c N_a + h.c.
\label{L}
\eeq

The neutrino mass eigenstates $\nu^{\rm (m)}_i$ ($i=1,...,n+3$) are linear
combinations of the weak eigenstates $\{\nu_\alpha, N_a \}$.  They are obtained
by diagonalizing the mass matrix:
\beq
\left( \begin{array}{cc}
          0 & y_{\alpha a} \langle H \rangle \\
y_{a \alpha} \langle H \rangle & {\rm diag}\{M_1,...,M_n\}
         \end{array}
\right)
\eeq
As long as all $y_{a \alpha} \langle H \rangle \ll M_a$, the eigenvalues of
this matrix split into two groups: the lighter states with masses of the order of
$y_{a \alpha} ^2 \langle H \rangle ^2/M_a$, and the heavier eigenstates with
masses of the order of $M_a$.  As usual, we will call the former {\em active
neutrinos} and the latter {\em sterile neutrinos}.  The mixing angles in this
case are of the order of $\theta_{a\alpha}^2 \sim y_{a \alpha} ^2 \langle H
\rangle ^2/M_a^2$.

The number $n$ of the right-handed singlets is unknown, although
it is clear that $n\ge 2$ is a necessary condition to explain the
results from the atmospheric and solar neutrino
experiments~\cite{2right-handed}. Theoretical considerations do
not constrain the number $n$ of sterile neutrinos. In particular, there is no
constraint based on the anomaly cancellation because the sterile fermions do
not couple to the gauge fields.  The experimental limits exist only for the
larger mixing angles~\cite{sterile_constraints}.  The scale of the
right-handed Majorana masses, $M_a$, can vary over many orders of magnitude. It
can be much greater than the electroweak scale~\cite{seesaw}, or
it may be as low as a few eV~\cite{deGouvea:2005er}.  It is also
possible that some of the right-handed Majorana masses are much
larger than others.  The seesaw mechanism can explain the
smallness of the neutrino masses even if the Yukawa
couplings are of order one, as long as the Majorana masses $M_a$ are large
enough.  However, the origin of the
Yukawa couplings remains unknown.  If the Yukawa couplings arise
as some topological intersection numbers in string theory, they
are generally expected to be of order one~\cite{Candelas:1987rx},
although very small couplings can are also
possible~\cite{Eyton-Williams:2005bg}.  However, if the Yukawa couplings
arise from the overlap of the wavefunctions of fermions located on
different branes in extra dimensions, they can be exponentially
suppressed and are expected to be very
small~\cite{Mirabelli:1999ks}.   If one or more singlets have
Majorana masses below the electroweak scale, they can appear as
sterile neutrinos and can have important ramifications; for
example, dark matter can be made up of sterile neutrinos with mass
of several keV~\cite{dw}, and the same particle can be responsible for the
observed pulsar kicks~\cite{pulsars}.

Several recent papers have studied in detail one particular case,
named $\nu$MSM~\cite{nuMSM}, which corresponds to $n=3$, $M_1\sim {\rm keV}$,
and $M_2\approx M_3\sim 1-10~ {\rm GeV}$.  In this model, the keV sterile
neutrino serves as the dark matter particle (and can explain the pulsar kicks),
while the degenerate heavier states, $M_2\approx M_3$, make the model
amenable to leptogenesis by neutrino oscillations~\cite{baryogenesis}.

The possible role of keV sterile neutrinos in astrophysics and cosmology, from
dark matter to pulsar kicks, to early star formation, makes the possibility of
their existence very intriguing.  However, if the neutrino Majorana masses
$M_a$ are below the electroweak scale, one should try to explain the origin of
this scale. The other fermions in the same mass range acquire their masses from
the Higgs mechanism.   Can the mass terms in eq.~(\ref{L}) also arise
from the Higgs mechanism?  The answer is yes; this requires an extension of
the Higgs sector by an SU(2) singlet field coupled to the righted-handed
fermions as in
Refs.~\cite{Chikashige:1980ht,Kusenko:2006rh,shaposhnikov_tkachev}:
\bea {\cal L}  &=&  {\cal L_{\rm SM}} + i \bar N_a
\slashed{\partial} N_a - y_{\alpha a} H^{\dag} \,  \bar L_\alpha N_a -
\frac{f_a}{2} S \; \bar N_a^c N_a   \nonumber
\\ &-& V(H,S) + h.c. \label{LwS}
\eea
We will assume that $S$ is a real scalar field to avoid the light
Nambu-Goldstone bosons associated with the breaking of the lepton number U(1);
the presence of such light bosons would render the sterile neutrinos unstable,
hence they could not be dark matter (although they could still explain the
pulsar kicks~\cite{pulsars}). If the singlet has very small mass and a
large VEV, it can be the inflaton~\cite{shaposhnikov_tkachev}.  We will not
discuss this interesting possibility here, but we will concentrate instead on a
singlet Higgs whose mass and VEV are both of the order of 100~GeV, which,
incidentally, is the requirement for the keV dark matter, as long as the mass
and VEV of $S$ are of the same order of magnitude~\cite{Kusenko:2006rh}.

As soon as the $SNN$  coupling is introduced in the
lagrangian, there appears a new way in which the relic population
of sterile neutrinos can be produced, namely from the decays
$S\rightarrow NN$.  This decay mechanism can
operate in addition to the neutrino oscillations mechanism of
Dodelson and Widrow~\cite{dw}, and one has to compare the relative
amounts produced by each of them.  Another important issue is how
cold the dark matter is if it is produced predominantly from the Higgs
decays.  Since the production occurs mainly at temperatures of the
order of the Higgs mass, $T\sim 100$~GeV, the reduction in the
number of degrees of freedom and the entropy production that takes
place as the universe cools down from $T\sim 100$~GeV causes the
dark matter population to be diluted and red shifted by a
factor $\xi \ge 33$ in the density and factor $\xi^{1/3} \ge 3.2$
in the average momentum.  These values reflect only the Standard Model degrees
of freedom, and any additional new physics will make $\xi$ even larger.   The
corresponding free-streaming length is shorter, and the Lyman-alpha bounds
become proportionately weaker~\cite{Kusenko:2006rh}.

In this paper we discuss the details of sterile dark matter production in a
model represented by the lagrangian (\ref{LwS}), with the scalar
potential
\bea V(H,S) &=& -\mu_H^2 |H|^2 - \frac{1}{2}\mu_S^2 S^2
+ \frac{1}{6}\alpha S^3 + \omega |H|^2 S  \nonumber
\\ &+& \lambda_H |H|^4 + \frac{1}{4}\lambda_S S^4 + 2\lambda_{HS}|H|^2 S^2
\label{V} \eea

\section{Sterile dark matter: cold or warm?}

If dark matter has a non-zero free-streaming length, the structure on small
scales may be suppressed. Studies of small-scale structure based on the
observations of dwarf Spheroids~\cite{dSphs} or Lyman-$\alpha$ forest
data~\cite{viel} can constrain or measure the free-streaming length of the
dark-matter particles, but the relation between this length and the particle
mass depends on the production mechanism.  One can
approximately relate the free-streaming
length to the mass $m_s$ and the average momentum of the sterile neutrino:
\begin{equation}
 \Lambda_{_{FS}} \approx 1.2 \, {\rm Mpc} \left ( \frac{\rm keV}{m_s} \right)
\left(  \frac{\langle p_s \rangle}{3.15 \, T}  \right)_{T\approx {\rm 1 keV}}
\label{free_streaming_length}
\end{equation}
This is a relatively good measure in many cases, although in general one has to
calculate the full power spectrum.  The observations of Lyman-$\alpha$ forest
constrain the free-streaming length to be less than 0.11~Mpc~\cite{viel}.  This
bound does not translate directly into a constraint on the mass because the
average momentum depends on the production mechanism.   For three scenarios
usually discussed in the literature,
\beq
\left(  \frac{\langle p_s \rangle}{3.15 \, T}  \right)_{T\approx {\rm
 keV}} =
\left \{
\begin{array}{cl}
0.8-0.9, &  {\rm for\  DW} \, \\
\approx 0.6, &  {\rm for} \  L\neq 0, {\rm  resonance }\, \\
\lesssim 0.2, & {\rm for}\ T_{\rm prod} \gtrsim 100 \, {\rm GeV} \,
\end{array}
 \right.
\label{red_shifting_factor}
\eeq
Here DW stands for Dodelson-Widrow production mechanism via non-resonant
neutrino oscillations~\cite{dw}, ``$L\neq 0$'' refers to the Shi--Fuller
production via the resonant neutrino oscillations in the case when the lepton
asymmetry is relatively large~\cite{shi_fuller}, and ``$ T_{\rm prod} \gtrsim
100~{\rm GeV}$'' refers to the production of sterile neutrinos at a temperature
well above the QCD scale, in which case the cooling and reduction of the
degrees of freedom causes the red shift in the population of dark
matter~\cite{Kusenko:2006rh}.

For the same mass, the sterile dark matter can be colder or warmer, depending
on the production mechanism.  This is clear from
equations (\ref{free_streaming_length}) and (\ref{red_shifting_factor}),
which, for a given cosmological scenario, relate the free-streaming length with
the mass.
Therefore, we will pay close attention to the factors that can
affect the momentum distribution in each scenario.

There are several ways in which the population of dark matter particles could
have formed in our model:
\begin{itemize}
\item The bulk of sterile neutrinos could be produced from neutrino
oscillations.  If the lepton asymmetry is negligible, this scenario~\cite{dw}
appears to be in conflict with a combination of the X-ray bounds~\cite{x-rays}
and the Lyman-$\alpha$ bounds~\cite{silk}, although it is possible to evade this
constraint if the lepton asymmetry of the universe is greater than
$O(10^{-3})$~\cite{shi_fuller}.  It is possible that the decays of additional,
heavier sterile neutrinos, can introduce some additional entropy and
contribute to cooling of dark matter~\cite{Asaka:2006ek}.  It is also possible
that the sterile neutrinos make up only a fraction of dark
matter~\cite{Kusenko:2006rh,silk}, in which case they can still be responsible
for the observed velocities of pulsars.

 \item The bulk of sterile neutrinos could be produced from decays of $S$
bosons at temperatures of the order of the $S$ boson mass, $T\sim 100$~GeV.
This scenario was discussed in Ref.~\cite{Kusenko:2006rh}.  In this case,  the
Lyman-$\alpha$ bounds on the sterile neutrino mass are considerably weaker
 than in the former case.

\item The decays described above could happen before a first-order phase
transition, and the entropy release in the transition could
redshift the population of the dark-matter particles.  We have
explored this possibility in detail, as discussed below, but we
have not found a range of parameters in which the phase transition
could  cool down the sterile dark matter significantly.

\item $S$ bosons could be so weakly coupled to the rest of the Higgs sector
that they would go out of equilibrium and decay
out of equilibrium at some temperature $T<100$~GeV.   As discussed below,
this scenario can produce a sufficient amount of dark matter.

\end{itemize}

We will now discuss these possibilities in detail.

\section{Production from the Higgs decays in equilibrium}

\begin{figure}[ht!]
  \centering
  \includegraphics[width=8cm]{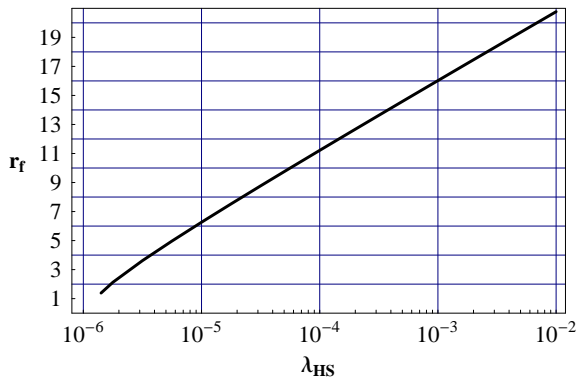}
  \caption{The variation of the $S$ boson freeze-out parameter $r_{\rm f}=m_S/T_{\rm f}$
   with the coupling to SM particles $\lambda_{HS}$. For numerical estimations
we used  $m_S=200\GeV$.}
  \label{lambda-rf}
\end{figure}

The interactions of the singlet Higgs bosons with SM particles have been studied
by McDonald in Ref.~\cite{McDonald:1993ex}, where the $S$ bosons were made
stable by imposing a global U(1) symmetry, which removed the odd power couplings,
and by setting $\mu_{_S}^2<0$, which forced $\langle S\rangle=0$. In this case,
the coupling $\lambda_{HS} $ controls the
$SS \rightarrow XX$ annihilations, into SM fermions and the $W,Z$ bosons.
We do not require $S$ to be stable. After $S$ develops a VEV,
other couplings also contribute to the
annihilations into SM particles. For each of these processes the cross section
for annihilation is:
\beq \sigma_{\rm ann} \sim 10^{-2} \frac{\lambda_{HS}^2}{m_S^2} \label{sigma} \eeq
At some temperature, these processes fail to
keep the $S$ particles in equilibrium, and they freeze out at $T_{\rm f} = m_S/r_{\rm f}$.
For very small $\lambda_{HS}\lesssim 10^{-6}$, S bosons never come into equilibrium.
A more detailed numerical calculation yields the dependence of the freeze-out
time
parameter $r_{\rm f}$ on $\lambda_{HS}$ shown in fig.~\ref{lambda-rf}.

The cubic couplings contribute to the annihilation processes through
exchange of virtual $S$ bosons. In fact, this will be the dominant process that
keeps $S$ particles in equilibrium, as long as $\frac{\alpha \omega}{m_S^2}
\gtrsim \lambda_{HS}$,
where $m_S$ is the $S$ boson mass. Comparing with Fig.~\ref{lambda-rf}, one can
see that this process can keep $S$ bosons in equilibrium down to rather low
temperatures, even if $\alpha, \omega$ are well below the $S$ mass.

Let us now assume that the $\alpha, \omega$ and/or $\lambda_{HS}$
are large enough (exact limit for $\lambda_{HS}$ to be defined below) to keep
$S$ in equilibrium down to temperatures well below its mass.
One can make a rough estimate of the sterile neutrino production by multiplying the $S$
number density by the $S \rightarrow NN$ decay rate $\Gamma = f^2
m_S/(16\pi)$ and by the time available for the decay, $\tau \sim
M_0/2T^2$, at the latest temperature at which the thermal population of $S$
is still significant, namely $T\sim m_S$. At lower temperatures,
the $S$ number density is too small, much smaller than $T^3$. One
obtains an approximate result
\beq
 \left( \left. \frac{N_s}{T^3}  \right) \right |_{T\sim m_S}\sim \Gamma \left.
\frac{ M_0}{T^2}\right |_{T\sim m_S} \sim \frac{f^2}{16\pi}
\frac{M_0}{m_S},
\eeq
where $M_0 = \left(\frac{45 M_{PL}^2}{4 \pi^3 g_*}\right)^{1/2}  \sim 10^{18} \GeV$
is the reduced Planck Mass.

This simple estimate is in agreement with the solution of the
kinetic equation discussed below. Of course, this description breaks
down if the $S$ particles decouple and
decay at a much later time. We will come back to this possibility.  For now,
let us assume that $S$ particles maintain their equilibrium populations
down to temperatures at least a factor of a few below their masses.   The dark
matter abundance for sterile neutrinos from the decay of bosonic particles in
equilibrium was first computed for a model in which the $S$ field served as the
inflaton with a potential adjusted to have $\langle S \rangle \gg
m_S$~\cite{shaposhnikov_tkachev}.  The results of this computation carry over
to our case.  Here we do not require $S$ to be the inflaton, and we take
$\langle S \rangle \sim m_S$, as in Ref.~\cite{Kusenko:2006rh}.  As was shown
in Ref.~\cite{Kusenko:2006rh}, the choice $\langle S \rangle \sim m_S$, along
with the requirement that sterile neutrinos make up all the dark matter, force
the $S$ boson mass and VEV to be right at the electroweak scale, $\langle S
\rangle \sim m_S\sim 10^2\, {\rm GeV}$, suggesting that $S$ may, indeed, be a
part of the Higgs sector of the extended Standard Model, and justifying some of
our assumptions regarding the Higgs potential.

Dark matter production from particle decays has been considered in a number of
papers~\cite{shaposhnikov_tkachev,decays}.  Let us first consider the decaying
particle in equilibrium.  As in the case of the inflaton
decay~\cite{shaposhnikov_tkachev}, the sterile neutrino distribution function
$n(p,t)$ is found from the following kinetic equation:
\beq \frac{\partial n}{\partial t} - H p \frac{\partial
n}{\partial p} = \frac{2 m_S \Gamma}{p^2}
\int_{p+\frac{m_S^2}{4p}}^{\infty} n_S dE,  \label{kinetic} \eeq
where $\Gamma = m_S f^2 /16\pi$ is the partial width of the $S$
boson decay. It is assumed that the
sterile neutrinos are never in equilibrium, and the inverse decays $N N
\rightarrow S$ can be neglected, which is true for small
Yukawa couplings $f< 10^{-7}$. Transforming to the variables: $r
= \frac{m_S}{T}$, $x=\frac{p}{T}$, one can rewrite eq. (\ref{kinetic}):
\beq \frac{\partial n}{\partial r} = \frac{f^2}{8 \pi}
\frac{M_0}{m_S} \frac{r^2}{x^2} \int_{x+\frac{r^2}{4x}}^{\infty}
\left.n_S\right|_{_{\frac{E}{T}=\zeta}} d\zeta .
\label{kinetic-tr}
\eeq

Since $S$ and $H$ mix, one has to consider the mixed mass
eigenstates in plasma. Both of them can decay into
sterile neutrinos. The SM Higgs is in thermal equilibrium due to
the coupling with the SM particles. For the temperature range in
which $S$ is also in equilibrium, the distribution
functions of the two mass eigenstates will be:
\beq n_j = \frac{1}{e^{E_j/T}-1} \eeq
Then eq.~(\ref{kinetic-tr}) yields:
\bea n^{\rm \Theta}(x,r) &=&
\sum_{j=1}^2  \frac{f_j^2}{8 \pi} \frac{M_0}{m_j}
\left[\frac{r_j^3}{3x^2} \ln \left(1-
e^{-x-\frac{r_j^2}{4x}}\right)^{-1} \right. \nonumber \\
&+& \left. \frac{8 x^2}{3} \int_1^{1+
\frac{r_j^2}{4x^2}} \frac{(z-1)^{3/2} dz}{e^{xz}-1} \right]
\label{equil-nsol} \eea
where the subscript $j=1,2$ runs over the two Higgs
mass eigenstates and the superscript $\Theta$ denotes production from decays
of $S$ bosons in equilibrium.  (We will use  $\displaystyle{\not} \Theta$ for
the case of $S$ bosons decaying out of equilibrium.)  In (\ref{equil-nsol}) the
first term is important when $r \lesssim 1$, while the second is
the dominant one for $r\gtrsim 1$. The above solution was obtained
assuming $f_j, \: m_j$ and the number of degrees of freedom $g_*$
remain constant. This is not valid  after the electroweak phase
transition takes place, since the Higgs mass eigenvalues and their
mixing are different in the two vacua. If $f_j, m_j$ or $g_*$ change
at some points in the evolution of the universe, the solution has to
be adjusted to include the contributions from all the periods corresponding
to different $f_j, m_j, g_*$. Each of these contributions will still be
given by (\ref{equil-nsol}), for the appropriate values of the
parameters and taken over the respective time intervals.

This  complication turns out to be irrelevant, since
the production rate of sterile neutrinos through each mode $m_j$
exhibits a peak at $r_j\simeq2.3$, which defines
the production temperature $T_{\rm prod}=m_j/2.3$. Most of the
sterile neutrinos are produced around that temperature, and by the
time when $r_j\approx10$ the production of
sterile neutrinos though $m_j$ decays has practically been
completed. More specifically, in the simplified case $m_1\ll m_2$,
$f_1\gg f_2$, $m_1$ decays will dominate over $m_2$ decays and the
total abundance of sterile neutrinos $Y_s = N_s/s$ at any later
temperature will be:
\beq Y^{\rm \Theta}_s (r) = \frac{45}{32 \pi^5}
\frac{f^2}{g_*(T_{\rm prod})} \frac{M_0}{m} \; y(r)
\label{Ys-equil} \eeq
where
\bea y(r) &=& \frac{1}{3} \int_0^\infty dx \left[r^3 \ln \left(1-
e^{-x-\frac{r^2}{4x}}\right)^{-1} \right. \nonumber
\\ &+& \left. 8 x^4 \int_1^{1 + \frac{r^2}{4x^2}} \frac{(z-1)^{3/2} dz}{e^{xz}-1}\right] \label{y-equil}
\eea
and we have dropped the mass eigenstate index for simplicity. The
$g_*(T_{\rm prod})^{-1}$ factor in (\ref{Ys-equil}) designates the fact that
the sterile neutrino population will be diluted by
\beq \xi =g_*(T_{\rm prod})/g_*(0.1\, {\rm MeV}) \label{xi}\eeq
as the universe cools down, due to the entropy release as the
effective degrees of freedom decrease.
At $r \rightarrow \infty$, the sterile neutrino abundance produced
from in-equilibrium decays takes the limiting value:
\beq Y_s^{\rm \Theta}(\infty) = \frac{27 \zeta(5)}{32 \pi^4}
\frac{f^2}{g_*(T_{\rm prod})}  \frac{M_0}{m} .\eeq

We require that the decays of $S$ bosons occur in equilibrium,
that is $T_{\rm f} \lesssim \frac{m}{10}$.  Then we get from
Fig.~\ref{lambda-rf}, $\lambda_{HS} \gtrsim  5 \cdot
10^{-5}$, for $\alpha = \omega =0$. However, if
$\alpha, \omega \gtrsim 1\, \GeV$ $S$ bosons stay in equilibrium
down to the desired temperature, regardless of $\lambda_{HS}$.

The momentum distribution in eq.~(\ref{equil-nsol}) is
non-thermal. Taking into account only the dominant decay
mode, one obtains (same as in Ref.~\cite{shaposhnikov_tkachev}) the momentum
distribution of dark-matter particles at $r \rightarrow \infty$:
\beq n^{\Theta}(x) =\frac{f^2}{3 \pi} \frac{M_0}{m}
x^2 \int_1^{\infty} \frac{(z-1)^{3/2} dz}{e^{xz}-1}, \eeq
for which the average momentum at temperature $T\sim 10^2$~GeV,
immediately after their production, is~({\em
cf.}~Ref.~\cite{shaposhnikov_tkachev})
\beq \left ( \frac{\langle p \rangle}{T} \right)_{T\sim 100\, {\rm
GeV}} = \frac{\pi^6}{378 \, \zeta(5)} \simeq 2.45. \eeq
This is lower than the same quantity for a thermal distribution, ${\langle
p \rangle}/{T} =3.15$.

Even more importantly, these momenta are further redshifted as the
universe cools down from the temperature at which most dark matter
is produced, $T_{\rm prod} \sim m \sim 100 \, {\rm GeV}$  to the much lower
temperatures at which the structure begins to form. As
the universe cools down, the number of effective degrees of
freedom decreases from $g_*(T_{\rm prod})=110.5$ to $g_*(0.1\,
{\rm MeV})=3.36$. This assumes no new physics below the Higgs mass;
any new physics would cause an additional cooling of the dark
matter. The ratio of dark matter to entropy is reduced by the
factor $\xi \approx 33$. This causes the redshifting of $\langle
p_s \rangle$ by the factor $\xi^{1/3}$:
\beq
\left ( \frac{\langle p \rangle}{T} \right)_{(T \ll 1{\rm MeV})} = 0.76 \,
\left [ \frac{110.5}{g_*(\tilde{m}_{_{S}}) }\right ]^{1/3}.
\label{p_s_redshifted}
\eeq
This is very different from the DW scenario~\cite{dw}, in which the
average neutrino momentum at low temperature $T$ is
\beq
\langle p_s \rangle_{\rm DW} = 2.83 \, T.
\label{p_s_DW}
\eeq
Comparing eqns. (\ref{p_s_DW}) and (\ref{p_s_redshifted}), one concludes that
the sterile neutrino mass corresponding to the same free-streaming length can
vary by more than a factor of 3 depending on the production
scenario~\cite{Kusenko:2006rh}.  A detailed analysis of the free-streaming properties of
``chilled'' dark matter is presented in Ref.~\cite{Petraki:2008ef}.

The dark matter abundance in this model depends on the details of
the Higgs mass matrix and the two-component decays, aside from
which it has the form:
\beq
\Omega_{\nu_s} \sim 0.2 \left( \frac{f}{10^{-8}}\right)^3  \left(
\frac{\langle S \rangle }{m_{1,2}}\right) \left( \frac{33}{\xi} \right).
 \label{omega} \eeq
Since we expect the masses of the two mass eigenstates to be of the same order,
and considering the cubic power of the unknown coupling $f$, the details of the
solution are not very important.  However, what may be important is the
additional effect of the first-order phase transition on the average momentum
of the dark matter particles.  If the dark matter population is redshifted
significantly by the entropy release in the phase transition, then the
Lyman-$\alpha$ bound could be further relaxed.

\section{Electroweak phase transition}

As discussed above, the population of sterile neutrinos  is subject to
dilution and redshift due to the entropy production that occurs (i) in any
possible  phase transitions at lower temperatures, and (ii) when the number of
degrees of freedom in plasma decreases due to the decoupling of Standard Model
particles below 100~GeV.
The redshift due to (ii) alone can reduce the momenta of dark
matter particles by a factor more than 3.2~\cite{Kusenko:2006rh}. Of
course, the dilution due to (i) matters only if a first-order
phase transition takes place after the sterile neutrinos are
produced, and, moreover, if $S$ bosons are too heavy to have a
high number density in the new vacuum after the phase transition.
To study this possibility, one has to take into account the
temperature effects on the effective potential and the history of
the phase transitions predicted by the model.  Electroweak phase transition in
a model with a singlet Higgs has been analyzed in
Refs.~\cite{Ahriche:2007jp,McDonald:1993ey_bgen,Profumo:2007wc}.  The
plausibility of the first-order phase transition makes the electroweak
baryogenesis a viable possibility, and it has implications for the LHC and the
ILC~\cite{Profumo:2007wc}.  Here we concentrate on the effects
the first-order transition could have on the population of dark-matter sterile
neutrinos.

\subsection{Finite-temperature effects and the first-order phase transition}

The tree level effective potential in terms of the VEV of the two Higgs bosons,
$\langle H \rangle = \frac{1}{\sqrt{2}} \eta$ and  $\langle S \rangle =\sigma$ is
\bea V^0_{\rm tree}(\eta,\sigma) &=&  -\frac{1}{2}\mu_{_H}^2 \eta^2 +
\frac{\lambda_{_H}}{4} \eta^4 - \frac{1}{2} \mu_{_S}^2 \sigma^2 +
\frac{\lambda_{_S}}{4} \sigma^4 \nonumber
\\ &+&  \lambda_{_{HS}} \eta^2
\sigma^2 + \frac{\alpha}{6} \sigma^3 + \frac{\omega}{2} \eta^2
\sigma \label{V0} \eea

To study the phase transition, we included the one-loop temperature-dependent
corrections and analyzed the potential numerically, as discussed below.

The tree level Higgs mass eigenvalues are:
%
\bea
(m^0_{_{1,2}})^2 &=& \frac{1}{2}\left[
\left(3\lambda_{_H} + 2\lambda_{_{HS}} \right) \eta^2 + \left(3\lambda_{_S} + 2 \lambda_{_{HS}}\right) \sigma^2  \right. \nonumber
\\ &+& \left(\omega + \alpha\right) \sigma -\mu_{_H}^2 -\mu_{_S}^2   \nonumber
\\ &\pm& \left\{ \left[ \left( 3\lambda_{_H} - 2\lambda_{_{HS}} \right) \eta^2
- \left(3\lambda_{_S} - 2\lambda_{_{HS}}\right) \sigma^2 \right. \right. \nonumber
\\ &+& \left. \left(\omega - \alpha\right) \sigma - \mu_{_H}^2 + \mu_{_S}^2  \right]^2 \nonumber
\\ &+&  \left. \left. 4 \eta^2 \left(4 \lambda_{_{HS}} \sigma + \omega  \right)^2 \right\}^{1/2} \right]
\label{mHS}
\eea
The 1-loop, zero temperature correction to the effective potential
\cite{Coleman:1973jx}
in the $\overline{MS}$ renormalization scheme is:
\beq
V^1(\eta,\sigma) = \sum_i \frac{n_i}{64 \pi^2} m_i^4(\eta,\sigma)
\left(\log \frac{m_i(\eta,\sigma)^2}{m_i(\eta_0,\sigma_0)^2} -\frac{3}{2}
\right)
\label{V1}
\eeq
where $n_i$ are the degrees of freedom of the contributing particles
and $m_i(\eta, \sigma)$ are their field-dependent masses.
The main contributions are from the neutral component of the
SM Higgs, the singlet Higgs, the Goldstone bosons
$\chi$, the $W$ and $Z$ gauge bosons and the top quark $t$:
\beq n_{_t}=-12, \; n_{_W}=6, \; n_{_Z}=3, \; n_{_\chi}=3, \; n_{_H} = n_{_S} = 1 \eeq
The gauge-singlet Higgs $S$ does not couple to the fermions or the gauge
bosons, thus their field dependent masses are the same as in the
minimal $SM$:
\bea m_{_t}^2 &=& \frac{y_t^2}{2}\eta^2 , \; m_{_W}^2 =
\frac{g^2}{4}\eta^2, \; m_{_Z}^2=\frac{g^2+g'^2}{4}\eta^2, \nonumber
\\ m_{_\chi}^2 &=& \lambda_{_H} \eta^2 - \mu_{_H}^2 + 2\lambda_{_{HS}} \sigma^2 + \omega \sigma \label{masses}
\eea
The Higgs mass eigenvalues are given by (\ref{mHS}). $m_i(\eta_0,\sigma_0)$ stand
for particle masses at the vacuum state $(\eta_0,\sigma_0)$ at zero temperature.

The temperature-dependent contribution to the effective potential at
one loop is \cite{Dolan:1973qd, Carrington:1991hz}:
\bea
 V^T(\eta, \sigma, T) = \sum_i \frac{n_i T^4}{2 \pi^2} I_{b,f}
\left(\frac{m_i^2(\eta,\sigma)}{T^2}\right) + V^T_{_{\rm ring}}
\label{VT}
\eea
where
\beq
I_{b,f}(y) = \int_0^\infty dx x^2 \log\left[1 \mp
e^{-\sqrt{x^2+y}} \right] \label{Ibf} \eeq
The upper sign corresponds to bosons, while the lower one to
fermions. Since we consider a wide range of temperatures,
we do not make use of the well known high-T expansion of the
functions (\ref{Ibf}).  $V^T_{\rm ring}$ is the ring contribution
of the gauge, Higgs and goldstone bosons:
\bea
V^T_{_{\rm ring}} &=& -\frac{T}{12 \pi} \left\{ {\rm Tr}\left[(m_{gb}^2 + \Pi_{gb})^{3/2} - (m_{gb}^2)^{3/2}\right] \right. \nonumber
\\ &+& {\rm Tr}\left[(m_{\rm higgs}^2 + \Pi_{\rm higgs})^{3/2} - (m_{\rm higgs}^2)^{3/2}\right] \nonumber
\\ &+& \left. n_{\chi}\left[(m_{\chi}^2 + \Pi_{\chi})^{3/2} - (m_{\chi}^2)^{3/2}\right] \right\}
\label{Vring}
\eea
where $m_{\rm higgs}$ is the tree level Higgs mass mixing matrix,
corresponding to the potential (\ref{V0}), whose eigenvalues are
given in (\ref{mHS}). $m_{\rm gb}^2$ is the mass mixing matrix for
the electroweak gauge bosons:
\bea
m_{\rm gb}^2 &=& \left(
                 \begin{array}{cccc}
                   \frac{g^2 \eta^2}{4} & 0 & 0 & 0 \\
                   0 & \frac{g^2 \eta^2}{4} & 0 & 0 \\
                   0 & 0 & \frac{g^2 \eta^2}{4} & -\frac{g g' \eta^2}{4} \\
                   0 & 0 & -\frac{g g' \eta^2}{4} & \frac{g'^2 \eta^2}{4} \\
                 \end{array}
               \right)
\label{mgb}
\eea
$\Pi_i$ are the thermal contributions to the masses, given for our model
by\cite{Carrington:1991hz,Ahriche:2007jp,Profumo:2007wc}:
\bea
\Pi_{\rm gb} &=& {\rm diag} \left[\frac{11}{6} g^2 T^2, \; \frac{11}{6} g^2 T^2,
 \; \frac{11}{6} g^2 T^2, \; \frac{11}{6} g'^2 T^2 \right] \nonumber
\\ \nonumber
\\
\Pi_{\rm higgs} &=&
{\rm diag}\left[ \left(\frac{3}{16} g^2 + \frac{1}{16} g'^2 + \frac{\lambda_{_H}}{2} + \frac{y_t}{4} + \frac{\lambda_{_{HS}}}{3} \right)  T^2, \right. \nonumber
\\ && \hspace{.9cm} \left. \left(\frac{1}{4} \lambda_{_S} + \frac{4}{3}\lambda_{_{HS}}\right) T^2 \right] \nonumber
\\ \nonumber
\\
\Pi_{\chi} &=&
\left(\frac{3}{16} g^2 + \frac{1}{16} g'^2 + \frac{\lambda_{_H}}{2}  + \frac{y_t}{4} + \frac{\lambda_{_{HS}}}{3} \right)  T^2
\label{thermal masses}
\eea

The effective potential at finite temperature is the sum of
(\ref{V0}), (\ref{V1}) and (\ref{VT}):
%
\beq
V_{\mathrm{eff}}(\eta,\sigma,T) = V^0(\eta,\sigma) +
V^1(\eta,\sigma) +V^T(\eta,\sigma,T) \label{Veff}  \eeq
In the calculations that follow we ignore $V^1$ for simplicity, since it
is only a small correction to the zero-T effective potential, and we treat the
imaginary part of the potential as usual~\cite{Weinberg:1987vp}.

\begin{figure}[ht!]
 \centering
 \includegraphics[width=6cm]{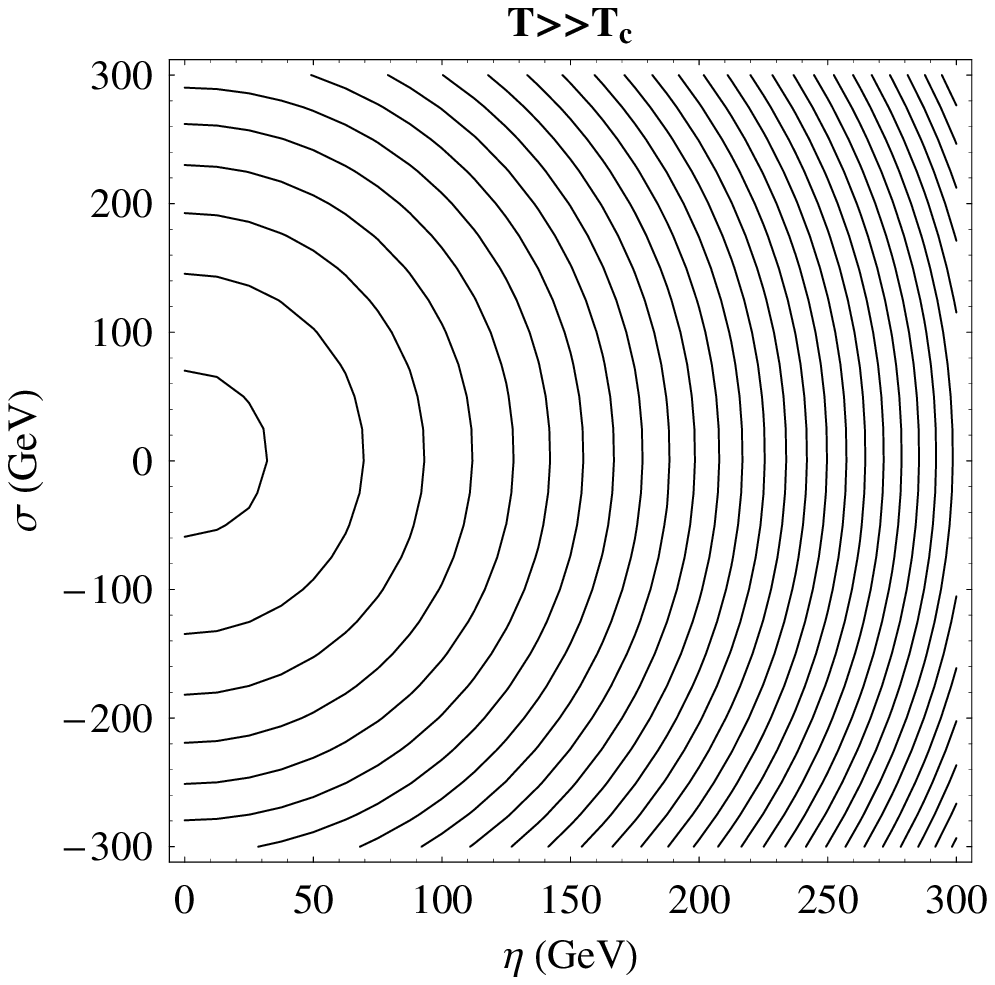}\\
 \includegraphics[width=6cm]{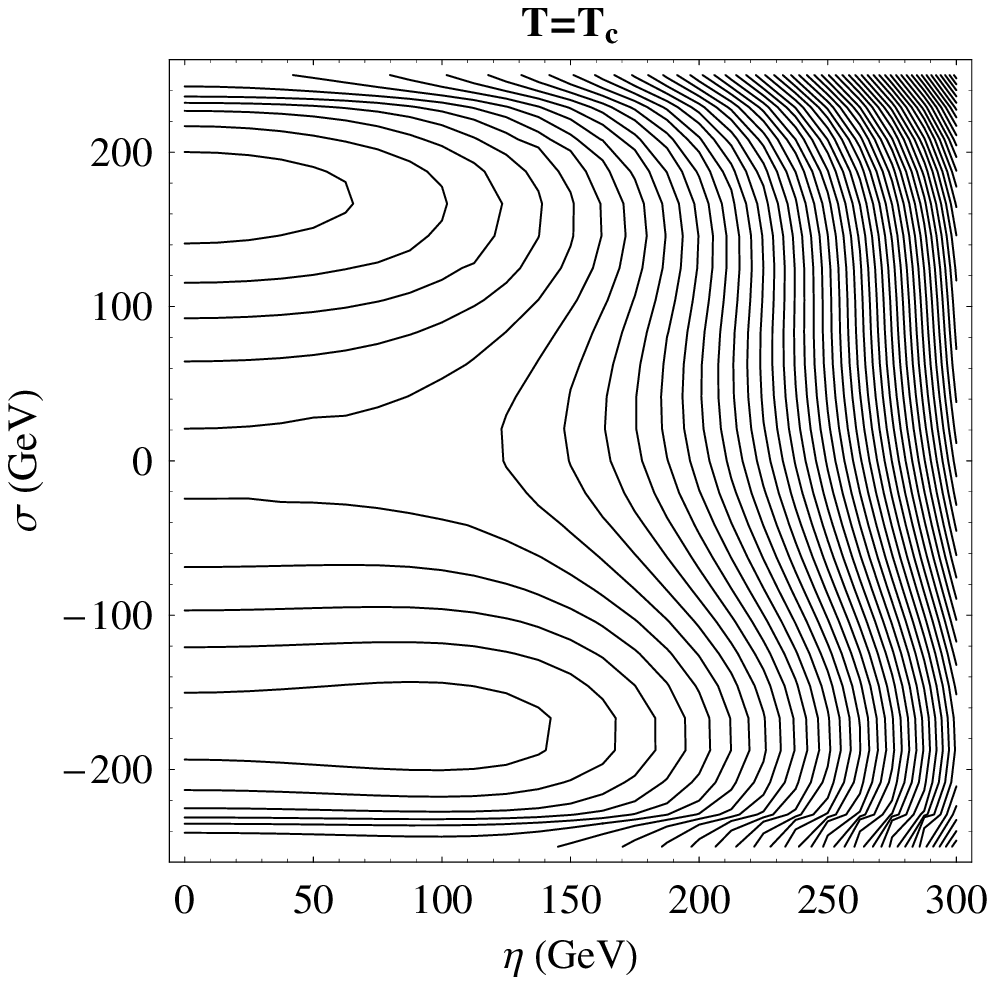}\\
 \includegraphics[width=6cm]{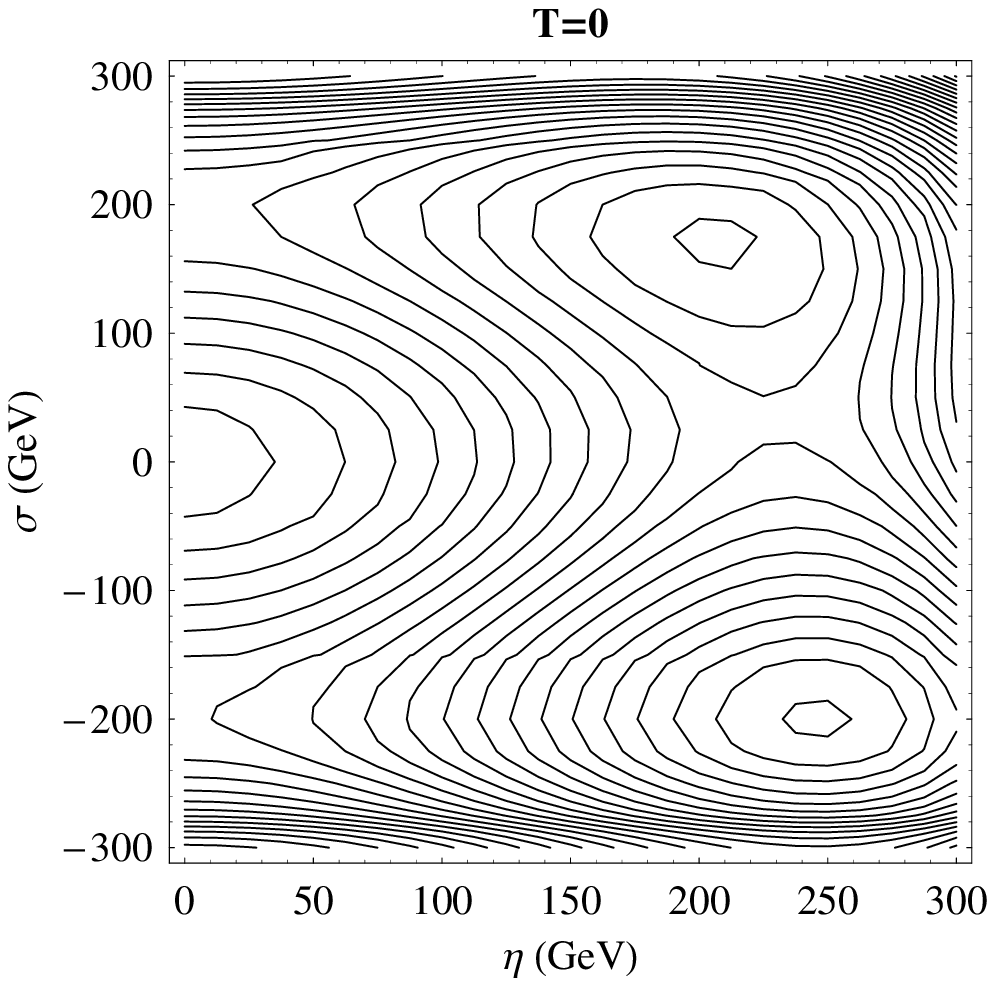}
 \caption{Contour plots of $V_{\mathrm{eff}}(\eta,\sigma,T)$ at $T\gg T_c$,
$T=T_c$, and  $T=0$, corresponding to the parameter set A of table
\ref{data}. At $T\gg T_c$ the universe is in the unbroken phase $\eta=0$, with
$\sigma>0$. At lower temperatures, the minimum of $V_{\rm eff}$ shifts to non-zero $\eta$,
while a second minimum appears in the $\sigma<0$ region. The two minima become
degenerate at $T=T_c$, and at $T=0$ the true vacuum is located at $\eta=246
\GeV$, $\sigma <0$.}
 \label{contours}
\end{figure}

The history of the universe for a typical set of parameters discussed in
table \ref{data}, is shown in fig.~\ref{contours}. Due to the
symmetry $\eta \rightarrow -\eta$, only the $\eta>0$ half plane
need to be considered.
Since the singlet does not couple to the gauge bosons and the $t$ quark, it
receives the smallest correction at a high temperature. Therefore, there is usually
a range of temperatures in which the Higgs doublet has no VEV, while
the singlet has a VEV. At a higher temperature, the singlet VEV
also tends to 0, although it never completely disappears due to the
$\sigma \rightarrow -\sigma$ asymmetry, induced by the $\alpha$ and
$\omega$ terms.

\begin{figure}[ht!]
 \centering
 \includegraphics[width=8cm]{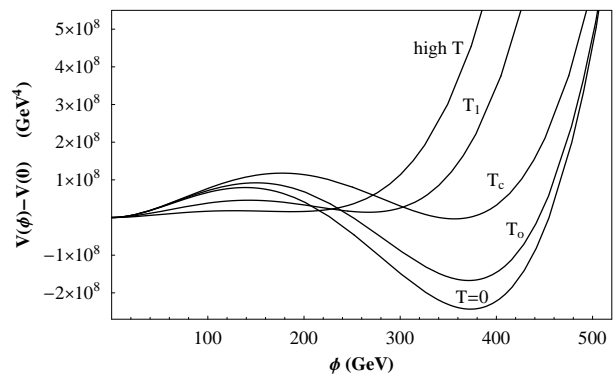}
 \caption{The potential configuration along the straight-line path connecting
the two minima, at various temperatures. At very high T the potential
possesses only one minimum. At a lower temperature $T_1$ a local minimum
starts forming. At $T_c<T_1$ the two minima become degenerate. At
$T_o<T_c$ tunneling to the true vacuum occurs. At $T=0$ the
universe has settled in the true vacuum. The curves correspond to parameter set
A of table \ref{data}.}
 \label{Vprofile}
\end{figure}

\begin{figure}[ht!]
 \centering
 \includegraphics[width=8cm]{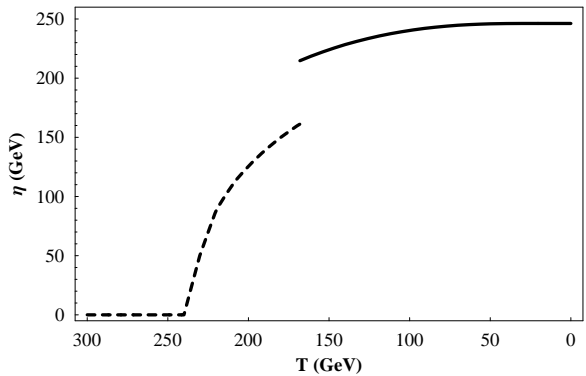}
 \caption{The evolution of the SM Higgs VEV $\eta$ with temperature. At very
high T, symmetry is restored: $\eta=0$. At a lower temperature a second order
phase transition to $\eta \neq 0$ takes place, while still remaining in the
false vacuum (dashed line). At $T=T_o$, a first order phase transition brings
the
universe to the true vacuum (solid line). At $T=0$, $\eta=246 \GeV$. The data
corresponds to parameter set A of table \ref{data}.}
 \label{Hvev}
\end{figure}

Therefore, in the early universe, at $T\gg 100$~GeV, the effective potential
has a unique minimum at $\{0,\sigma_{\rm f}\}$. At a lower temperature, for some
range of the parameter space, the doublet also develops a VEV and the universe shifts to
$\{\eta_{\rm f}(T),\sigma_{\rm f}(T)\}$ through a second order phase transition. 
In the meanwhile, a local minimum has been developed at 
$\{\eta_{\rm t}(T),\sigma_{\rm t}(T)\}$. At $T=T_c$ the two minima become
degenerate and at $T<T_c$, $\{\eta_{\rm t},\sigma_{\rm t}\}$ turns out to be
the true vacuum ({\em cf.}~Fig.~\ref{contours}). At some temperature
$T_o\lesssim T_c$ the false vacuum decays into the true vacuum via bubble nucleation,
described in section \ref{bubbles}. Fig.~\ref{Vprofile} shows
this evolution along the straight-line path connecting the two
minima, as universe cools down. In Fig.~(\ref{Hvev}), we show the
evolution of the order parameter, the SM Higgs VEV $\eta$.

\subsection{Phase Transition through Bubble Nucleation \label{bubbles}}

The tunneling from the false to the true vacuum was calculated numerically
using an approximation in which the bounce~\cite{Coleman:1977py} was assumed to
lie along a straight line in the 2-d field space ($\eta,\sigma$). Let $\phi$
be the field configuration along this path.
At finite temperature, one looks for solutions of the
Euclidean equations periodic in the ``time" direction with period
$T^{-1}$\cite{Linde:1981zj}.  In the high-temperature limit, the solution
should be translationally invariant along the ``time" axis, thus the
dependence of $\phi$ on temperature disappears. The
$O(3)$-symmetric (in the spatial coordinates) solution will now
obey the equations:
\beq \frac{d^2 \phi}{d r^2} +
\frac{2}{r} \frac{d \phi}{d r} = \frac{d V}{d \phi} ,
\hspace{.5cm} \left.\frac{d \phi}{dr}\right|_{r = 0} = 0,
\hspace{.5cm} \phi(\infty)=0,   \label{ELO3} \eeq
where $\phi=0$ is the false vacuum. The decay rate per unit volume is
\beq \Gamma \approx T^4 \left(\frac{S_3(T)}{2
\pi T}\right)^{3/2} e^{-S_3/T} \label{Gamma3},
\eeq
where we neglect the prefactor due to the change in the symmetry
group~\cite{Kusenko:1995bw}.  Here $S_3[\phi]$ is the 3-dimensional action:
\beq S_3[\phi] = \int d^3
x \left[\frac{1}{2}\left(\nabla \phi\right)^2 + V(\phi,T) \right]
\label{S3}. \eeq

For the solution of eq. (\ref{ELO3}) we adopt
numerical methods, rather than using the well-known approximation
schemes at the thin  and thick wall limits \cite{Linde:1981zj}.
The $O_4$ symmetric case at $T=0$ has been solved
numerically in \cite{Sarid:1998sn}. We do the same for the $O_3$
symmetric equation (\ref{ELO3}) and present here the results, used
for the numerical estimates of table \ref{data}.

A potential of the form: \beq V(\phi) = \frac{1}{2}M(T)^2 \phi^2
-\frac{1}{3} \delta (T) \phi^3 + \frac{1}{4} \zeta(T) \phi^4
\label{Vphi}\eeq with $M^2>0$ to ensure at least metastability at
$\phi=0$, encompasses all of the renormalizable potentials. For
such a potential the transition between different regions depends
on a single dimensionless parameter: \beq \kappa = \frac{9}{8}
\frac{\zeta M^2}{\delta^2} \label{kappa} \eeq For tunneling to
occur we must have $\kappa \leqslant \frac{1}{4}$, while $\kappa
\geqslant 0$ is required for the potential to be bounded from
below.

At finite temperature, the $O_3$ symmetric action (\ref{S3}) is
found to be:
\bea S_3 = \frac{9 M^3}{2^{3/2}
\delta^2} \times \left[\hat{S}_{\rm 3,thick} \right. &+& 125.8 \kappa - 239.5
\kappa^2  \nonumber \\
&+& \left. \frac{33 \kappa}{1-4\kappa} + \hat{S}_{\rm 3,thin} \right]
\label{S3num} \eea
with $\hat{S}_{\rm 3,thick} \simeq 13.72 (1-4 \kappa)^2$ and
$\hat{S}_{\rm 3,thin} = \frac{2^4 \sqrt{2} \pi}{3^6}
\frac{4\kappa}{(1-4\kappa)^2}$ being the limits for the thick and
thin wall approximation, respectively \cite{Linde:1981zj}.

\vspace{.5cm}

The time needed for the universe to tunnel from the false to the
true vacuum is estimated by setting $\Gamma \cdot t_H^4
\thickapprox 1$ where $t_H = \frac{M_0}{2T^2}$ is the Hubble time
in the radiation-dominated universe. Taking into account that during the
electroweak phase transition $T\thickapprox 100$ GeV, this yields:
$ \frac{S_3}{T} \thickapprox 160 \label{160}$, which defines the
tunneling temperature $T_o$ in the estimations of table
\ref{data}. If $\frac{S_3}{T} \ll 160$, the tunneling occurs very
quickly, when the universe is still hot. If $\frac{S_3}{T} \gg
160$, the tunneling rate is too low and the universe remains at
the false vacuum.

\begin{figure}[ht!]
  \centering
  \includegraphics[width=8cm]{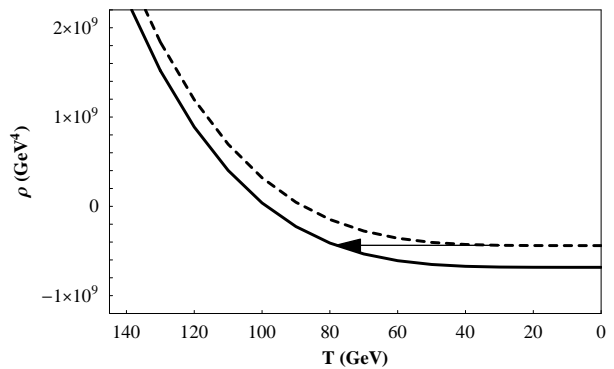}
  \caption{The free energy density \textsl{vs} temperature, at the true vacuum
(lower curve) and at the false vacuum (upper curve). Significant reheating
could occur if the universe cooled down to a low temperature before the
tunneling took place. At the temperature at which the tunneling actually
occurs, $T\approx 100 \GeV$, the reheating is not significant.}
  \label{rho_graph}
\end{figure}

When this condition is met, the universe tunnels from the false
vacuum $\phi_{\rm f}$ to the true vacuum $\phi_{\rm t}$. The
energy gained from the transition to a deeper minimum reheats the
universe from the tunneling temperature $T_o$ to a higher
temperature $T_r$. Since the expansion of the universe is much
slower than the tunneling, the reheating temperature $T_r$ is
found by taking the energy density to be constant $\rho (\phi_{\rm
f}(T_o), T_o) = \rho (\phi_{\rm t}(T_r), T_r)$, where $\rho
(\phi,T) = f(\phi,T) + T \; s(\phi,T)$, with $f(\phi,T) =
V_{\mathrm{eff}}(\phi, T)$ the free energy density and
$s=-\partial f/\partial T$ the entropy density. In fig.\ref{rho_graph} the free
energy density in the true and false vacuum are
presented vs temperature.

Representative numerical estimations done using the above are
presented in table \ref{data}.
\begin{table}[ht!]
  \centering
\begin{tabular}{|l|c|c|}
  \hline
  parameter sets & A & B   \\
  \hline
  $\lambda_H$        & 0.5          &  0.6       \\
  $\lambda_S$        & 0.6          &  0.4       \\
  $\lambda_{HS}$     & 0.025        & -0.02      \\
  $\alpha$           &    2         & -  25      \\
  $\omega$           &   25         &    90      \\
  $\sigma_0$         & -200         &  -220      \\
  \hline
  $T_c$              & 266          &   220      \\
  $T_o$              & 168          &   179      \\
  $T_r$              & 170          &   182      \\
  $v_c/T_c$          & 1.4          &   1.5      \\
  \hline
  $m_{1,2}^t(0)$     & 227, 247     & 188, 302   \\
  $m_{1,2}^f(T_o)$   & 191, 139     & 118, 101   \\
  $m_{1,2}^t(T_r)$   & 215, 201     & 251, 143   \\
  \hline
  $\langle p \rangle / T$   & 0.84  &  0.85      \\
  $f$                       & $4.2 \cdot 10^{-8}$  &  $4 \cdot 10^{-8}$   \\
  $m_s$ (keV)               & 8.42  &  8.85      \\
  \hline
\end{tabular}
 \caption{Representative parameter sets (see text for discussion).  The unit
for all dimensionfull parameters is GeV, except for $m_s$, which
is given in keV.}
\label{data}
\end{table}
The independent parameters of the model were chosen to be $\lambda_H,
\lambda_S, \lambda_{HS},
 \alpha, \omega$ and the VEVs of the two Higgs bosons at zero temperature
 $\eta_0, \sigma_0$, of which $\eta_0$ is kept fixed at 246~GeV.

$T_c, \: T_o, \: T_r$ stand for the critical, tunneling and
reheating temperature respectively. $v_c$ is the distance between
the two vacua at the critical temperature and $\frac{v_c}{T_c} >
1$ is the criterion for a strong 1st order phase transition. The
parameter space of the potential (\ref{V}) can provide for a variety
of phase transition scenarios (1st order only, 2nd order only,
2nd order followed by 1st order). The parameter sets
presented here fulfill the requirement for a
strong 1st order P.T. In A, a second order phase transition to
non-zero VEV of the SM Higgs precedes the first order one to the
true vacuum (fig.~\ref{Hvev}). In B, no second-order phase transition occurs.

$m_{1,2}^{\rm t}(T), m_{1,2}^{\rm f}(T)$ stand for the Higgs mass
eigenvalues at the true and false vacuum respectively, at temperature $T$.
For the parameter sets of table \ref{data}, both of the Higgs modes decay after
the transition
to the true vacuum, since $T_r > \frac{m_{1,2}^t(T_r)}{2.3}$,
which is the temperature at which the decay rate appears to be
maximal. Conversely, decay in the false vacuum would require
tunneling temperature small enough, $T_o <\frac{m^f(T_o)}{2.3}$,
in order for the Higgs bosons to have time to decay before the phase
transition,
and also sufficiently heavy Higgs eigenstates in the true vacuum,
$m^t(T_r) > 2.3~T_r $, so that decays after the phase transition are suppressed
by
the low number density of Higgs bosons. No parameter sets satisfying the above
were
found.

The values of the sterile neutrino Yukawa coupling $f$ to the
Higgs singlet, presented in table \ref{data}, are obtained by
requiring that sterile neutrinos make up all the dark
matter, where now the details of the two-component decay,
the phase transition and the decoupling of degrees of freedom were
taken into account. The numerical results are consistent with
the estimate of eq. (\ref{omega}). The sterile neutrino mass $m_s = f\cdot
\sigma_0$ is then
set to be in the keV range.

\section{Sterile neutrino production from out-of-equilibrium decays}
Finally, we address the possibility of $S$ decoupling early from
equilibrium  and decaying into sterile
neutrinos out of equilibrium. This is the case if
$\alpha, \omega \approx 0$ and $\lambda_{HS} \approx 10^{-6}$.
Then, only a second order, rather than a first order, phase transition takes place
and the $S$ decays occur in the broken phase.

The sterile neutrino population is again found from eq.
(\ref{kinetic-tr}), where now we need to first determine the
out-of-equilibrium concentration of $S$ bosons.

The $S$ boson number density $N_S$ after decoupling, taking into account
the annihilations of $S$ bosons to SM particles, is given
by~\cite{McDonald:1993ex}:
\beq \frac{N_S(T)}{T^3} =
\frac{N_S^{eq}(T_f)}{T_f^3}\frac{1}{2-r_{\rm f}/r}
\label{NS-McD-simpl} \eeq
If it were only for the $SS\rightarrow XX$ annihilations, the $S$
boson abundance  would decrease at $r\rightarrow \infty$ to just
half of its equilibrium value at freeze-out.  However,
the decay of $S$ particles to sterile neutrinos will result in an
exponential damping of the $S$ boson abundance. In addition, after $H$ and $S$
develop VEVs, $S$ bosons will decay to SM fermions through the mixing with the
SM Higgs. We can therefore ignore the $SS\rightarrow XX$ annihilations
and  consider only the $S\rightarrow N_a N_a$
and $S\rightarrow \bar{f} f$ decays to determine $n_S$ after
freeze out. The kinetic equation for $S$ bosons is:
\beq  E\frac{\partial n_S}{\partial t} - H|\vec{p}|^2
\frac{\partial n_S}{\partial E} = -\frac{m^2 h^2}{8\pi}
n_S \label{S kin}, \eeq
where
\bea
h^2 &\equiv& \sum_a f_a^2 \left(1-\frac{4f_a^2 \sigma^2}{m^2}\right) \nonumber
\\ &+& \sum_{\rm f} \lambda_f^2 \left(1-\frac{4m_f^2}{m^2}\right)
\left(\frac{\lambda_{HS}}{\max (\lambda_H,\lambda_S)}\right)^2
\label{h} \eea
takes into account the decay to all of the sterile neutrino
species and SM fermions
\footnote{We note in passing that sterile neutrinos with MeV$<M_a<M_S$
($a\ge 2$), produced in the $S$ decays, can decay into three active
neutrinos via mixing, in a tree-level process
that involves $Z$ exchange. Hence, we must take into account only the branching
ratio of decay into the long-lived singlets with with keV-scale masses.}.
Here $\lambda_f$ are the Yukawa couplings
of the SM fermions to the Higgs doublet and
$\frac{\lambda_{HS}}{\max (\lambda_H,\lambda_S)}$
is the mixing angle of the two Higgs mass eigenstates,
at the limit $\lambda_{HS}\ll \lambda_H,\lambda_S$ and
$\sigma \approx \eta$. Since $S$ bosons live in the electroweak scale,
the main fermion decay mode will be the $\bar{b} \: b$ channel.
For $\lambda_{HS} \gtrsim 10^{-6}$, this dominates over the decays
into keV sterile neutrinos. As we will see below,
$\lambda_{HS} \thickapprox 10^{-6}$ and $M_a \thicksim \keV$ is a
self-consistent set of parameters
for producing a sufficient amount of sterile neutrinos
to make up dark matter, through out-of-equilibrium decays
of $S$ bosons.

In terms of $r=m/T$ and $x_S=p_S/T$, one obtains:
\beq \frac{\partial n_S}{\partial r} = -  \frac{h^2
M_0}{8\pi m} \frac{r^2}{\sqrt{x_S^2+r^2}} n_S \label{S kin tr}.
\eeq
This yields:
\bea n_S(x_S,r) &=& \frac{1}{e^{\sqrt{x_S^2+r_{\rm f}^2}}-1}
\left(\frac{r+\sqrt{x_S^2+r^2}}{r_{\rm f}+\sqrt{x_S^2+r_{\rm
f}^2}}\right)^{\Lambda x_S^2} \times \nonumber \\
&& e^{-\Lambda (r \sqrt{x_S^2+r^2} - r_{\rm f} \sqrt{x_S^2+r_{\rm f}^2})}
\label{nS} \eea
where we set $\Lambda = \frac{h^2 M_0}{16 \pi m}$ and we
took $n_S$ to be the thermal equilibrium distribution function at
$T=T_{\rm f}$.

\begin{figure}[ht!]
  \centering
  \includegraphics[width=8cm]{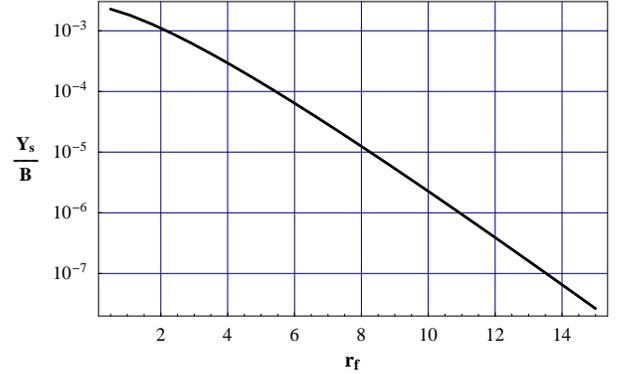}
  \caption{The final sterile neutrino abundance $Y_s^{^{\displaystyle{\not} \Theta}}(\infty)/B$ 
  vs $r_{_{\rm f}}$. $r_{_{\rm f}}\thicksim1-2$ corresponds to out-of-equilibrium decays of $S$ 
  bosons, producing sufficient amount of dark matter. For $r_{_{\rm f}}>3$, the amount produced is
  insignificant in comparison to that produced during the in-equilibrium decays, occurring at $r<3$, 
  and can be ignored.}
  \label{y final}
\end{figure}

Using (\ref{S kin tr}), eq. (\ref{kinetic-tr}) can be partially integrated to
give the
sterile neutrino distribution function, produced by $S$ bosons decays
after their freeze-out:
\bea n^{^{\displaystyle{\not} \Theta}}(x,r) &=& \frac{B}{x^2}
\left[ \int_{\left| \frac{r_{\rm f}^2}{4x}-x \right|}^\infty x_S n_S(x_S,r_{\rm
f}) dx_S  \right. \nonumber
    \\ &-&  \int_{r_{\rm f}}^r \frac{r'}{2x}\left(\frac{r'^2}{4x}-x\right)
n_S\left(\left|\frac{r'^2}{4x}-x\right|,r'\right)dr'    \nonumber
    \\ &-&  \left. \int_{\left| \frac{r^2}{4x}-x \right|}^\infty x_S n_S(x_S,r)
dx_S   \right]
\label{n-ooe-sol}   \eea
where we set $n^{{\displaystyle{\not} \Theta}}(x,r_{\rm f})=0$ and
\beq B \equiv \frac{f^2}{h^2} \label{B}\eeq
is the branching ratio of $S\rightarrow N_1 N_1$ to all other decays, with
$N_1$ being the lightest sterile neutrino.
The last term in (\ref{n-ooe-sol}) vanishes at the limit $r \rightarrow
\infty$, while the second term
does not contribute to the total abundance, but only shifts the momentum
distribution. The abundance of sterile neutrinos at any later time will be 
proportional to the amount of $S$ bosons that have already decayed up to that time:
\beq
Y_s^{^{\displaystyle{\not} \Theta}} (r) = B [Y_S(r_{\rm f}) -
Y_S(r)]
\label{Ys-ooe}\eeq
where
\beq Y_S(r) = \frac{45}{4 \pi^4 g_*} \int_0^\infty n_S(x_S,r)  x_S^2  dx_S
\label{YS} \eeq
is the $S$ boson abundance, with $n_S(x_S,r)$ given by (\ref{nS}).
The production rate $dY^{^{\displaystyle{\not} \Theta}}_s/dr$ peaks at some
$r_{\rm prod} = m/T_{\rm prod}$. Since $\Lambda$ determines how fast $S$ bosons decay,
$r_{\rm prod}$ depends on $\Lambda$ but is effectively independent of $r_{\rm
f}$. Given that $\lambda_{HS} \geqslant 10^{-6}$ is required for $S$ bosons to be in 
equilibrium at early times, $\Lambda$ receives a minimum contribution from the 
$\bar{b}b$ decay mode, through the mixing with the SM Higgs. Taking into account 
the $b$ quark Yukawa coupling to the SM Higgs $\lambda_b\simeq 2 \cdot 10^{-2}$ and that
$\lambda_H , \lambda_S <1$, the decay into $\bar{b}b$ pairs ensures that
$\Lambda \geqslant 0.01$, which results in $S$ bosons decaying early enough, 
at $r_{\rm prod}<10$, before the decoupling of the QCD degrees of freedom. 
Then, in eq.~(\ref{YS}), $g_* \thickapprox 90-110$, or $\xi\thickapprox 25-33$.
The dependance of the final sterile neutrino abundance 
$Y_s^{^{\displaystyle{\not} \Theta}}(\infty)/B$ on $r_{\rm f}$ is shown in Fig.{\ref{y final}}.

The amount of dark matter produced from the out-of-equilibrium decay of $S$
bosons is:
\beq \Omega_{\nu_s} \approx 0.2 \left( \frac{m_s}{3 \keV} \right) \left(
\frac{Y^{^{\displaystyle{\not} \Theta}}_s/B}{10^{-3}} \right)
\left(\frac{B}{0.1}\right)   \label{omega-ooe} \eeq
Early decoupling of $S$ bosons  $1 \lesssim r_{\rm f}\lesssim 3$ implies
$\lambda_{HS} \simeq 10^{-6}$ ({\em cf.}~Fig.~\ref{lambda-rf}). The dominant decay
mode is then $\bar{b}b$ pairs and the branching ratio of $S$ decays into an
$m_s\simeq 3 \keV$ sterile neutrino, i.e. with $f=(1-5) \times 10^{-8}$, is
$B\simeq 0.1-0.01$. The amount of sterile neutrinos produced by the 
out-of-equilibrium decays is then sufficient to constitute dark matter.

\begin{figure}[ht!]
  \centering
  \includegraphics[width=8cm]{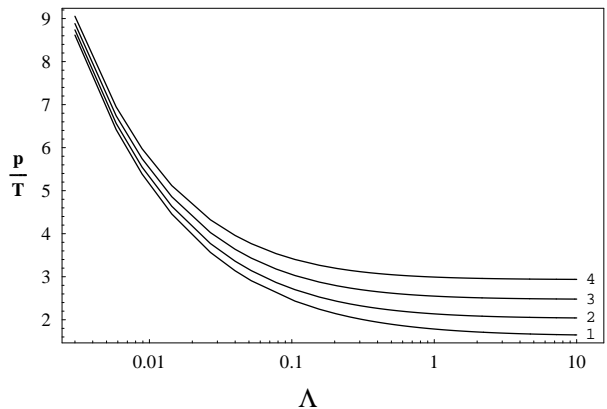}
  \caption{Final average momentum $\langle p \rangle$/T vs $\Lambda \propto h^2$, 
  for various values of $r_{\rm f}$, stated next to each curve. 
  $\langle p \rangle/T$ is further redshifted by $\xi^{1/3} \approx 3.2$ with 
  respect to the values presented in this graph. Small $\Lambda$ implies delayed decays, 
  resulting in warmer dark matter. In the out-of-equilibrium decay
  scenario, $\Lambda \thickapprox 0.1$ results in production of sufficient amount of dark matter.}
  \label{p final}
\end{figure}

The average momentum at $r\rightarrow \infty$ of the sterile neutrino
population produced through the out-of-equilibrium decays is
\beq \frac{\langle p \rangle}{T} = \frac{\Lambda \int_{r_{\rm f}}^\infty dr r^2
\int_0^\infty dx x^2 n_S(x,r)}
{\int_0^\infty dx x^2 n_S(x,r_{\rm f})} \label{p/T}.
\eeq
The variation of $(\langle p \rangle/T)$ with $\Lambda$ is shown in
Fig.~\ref{p final}, for various values of $r_{\rm f}$, where the redshifting
factor $\xi^{1/3}$ (eq.~(\ref{xi})) has not yet been included.
Then, for the set of parameters discussed above $\Lambda\approx 0.01 - 0.1$ and the
production peaks at $r_{\rm prod} \simeq 3$. Thus, the
average momentum at the current temperature, for decoupling around
$r_{\rm f} \sim 1-2$, can be as low as ({\em
cf.}~Fig.~\ref{p final}):
\beq \frac{\langle p \rangle}{T} \simeq  \left( \frac{2.5}{\xi^{1/3}} \right)_{_{T \ll 1 \MeV}} = 0.8
\label{p-ooe}. \eeq

\medskip

Finally, we return to the in-equilibrium decay scenario. If $r_{\rm f} > 3$,
$S$ bosons decay primarily while still in equilibrium. However, the amount
of dark matter produced will be supplemented by any additional sterile
neutrinos produced after $S$ bosons come out of equilibrium. This again is
given by eq.~(\ref{omega-ooe}), where now both $Y_s/B$ and $B$ take
lower values because of the increase in $r_{\rm f}$ and $\lambda_{HS}$ ({\em
cf.}~Fig.~\ref{y final}).
Thus, the amount of dark matter produced is determined only by a fraction of
what was produced before the freeze-out, and our results from the
previous section remain valid.

\section{Baryogenesis}

The model under consideration and its minimal modifications offer at least two
scenarios for generating the baryon asymmetry of the universe below the
electroweak scale.  One possibility is that the baryon asymmetry could arise
from the low-scale leptogenesis if there are at least three sterile neutrinos
 below the electroweak scale, and if the two heavier ones
are nearly degenerate in mass~\cite{baryogenesis}.  This scenario is different
from the more commonly discussed thermal leptogenesis in that neutrino
oscillations, not decays, are responsible for the change in the lepton number
of the plasma.  Active neutrinos (in equilibrium) can oscillate into the
sterile neutrinos (out of equilibrium), and CP violation in the neutrino mass
matrix could make the the net lepton number of the out-of-equilibrium sterile
neutrinos non-zero.  The excess lepton number remaining in plasma is partially
converted into the baryon number by sphalerons~\cite{baryogenesis}.

An alternative possibility exists if the phase transition is strongly
first-order, which is quite likely in the model with the singlet, as discussed
in Refs.~\cite{McDonald:1993ey_bgen,Ahriche:2007jp,Profumo:2007wc}.  In this
case the standard electroweak baryogenesis~\cite{krs} can take place in the
course of this first-order phase transition.  The model of eq.~(\ref{LwS}) can
easily be modified to include a sufficient amount of CP violation: all that is
required for a successful baryogenesis is to include the second Higgs
doublet~\cite{McDonald:1993ey_bgen}.

\section{Conclusion}

The inclusion of singlet fermions (right-handed neutrinos) in the Standard
Model is the usual way to generate the observed neutrino masses~\cite{seesaw}.
In contrast with many other models, we assume that both Dirac and Majorana
neutrino masses are generated via the Higgs
mechanism~\cite{Chikashige:1980ht}. The immediate advantage of this model is
the possibility to produce dark matter, in the form of sterile neutrinos, which
is cold enough to satisfy the bounds on the small-scale structure and the
bounds from X-ray observations, while explaining the pulsar kicks at the same
time~\cite{Kusenko:2006rh}. The same sterile neutrinos  can play
an important role in the formation of the first stars~\cite{reion}. We have
considered different ways in which the dark-matter sterile neutrinos
can be produced from the Higgs decays in the early universe.  If the
production from the Higgs decays dominates over the production by neutrino
oscillations, the resulting dark matter population is colder than in the
Dodelson-Widrow case for the same mass. The Higgs structure of the model has
important implications for the collider physics an can be probed at the
at the Large Hadron Collider and a Linear
Collider~\cite{Profumo:2007wc,singlet_higgs_LHC}.

This work was supported in part by the DOE grant DE-FG03-91ER40662 and by the
NASA ATP grant NAG~5-13399.  A.K. appreciates hospitality of the Aspen
Center for Physics.


\begin{thebibliography}{99}


\bibitem{seesaw} P. Minkowski, Phys. lett. {\bf B67 }, 421
(1977); M.~Gell-Mann, P.~Ramond, and R.~Slansky, \emph{Supergravity}
(P.~van Nieuwenhuizen et al. eds.), North Holland, Amsterdam, 1980;
T.~Yanagida, in \emph{Proceedings of the Workshop on the
Unified Theory and the Baryon Number in the Universe} (O.~Sawada
and A.~Sugamoto, eds.), KEK, Tsukuba, Japan, 1979; S.~L.
Glashow, in  \emph{Proceedings of the 1979 Carg{\`e}se Summer Institute on
Quarks and Leptons} (M.~L{\'e}vy et al. eds.), Plenum Press, New York,
1980; R.~N. Mohapatra and G.~Senjanovi{\'c}, Phys. Rev. Lett.
\textbf{44}, 912 (1980).

\bibitem{dw}
S.~Dodelson and L.~M.~Widrow,
Phys.\ Rev.\ Lett.\  {\bf 72}, 17 (1994).

\bibitem{production_oscillations}
K.~Abazajian, G.~M.~Fuller and M.~Patel,
Phys.\ Rev.\ D {\bf 64}, 023501 (2001);
%
A.~D.~Dolgov and S.~H.~Hansen,
Astropart.\ Phys.\  {\bf 16}, 339 (2002);
  K.~Abazajian,
  Phys.\ Rev.\ D {\bf 73}, 063506 (2006);
  T.~Asaka, M.~Laine and M.~Shaposhnikov,
  JHEP {\bf 0606}, 053 (2006);
  JHEP {\bf 0701}, 091 (2007);
  D.~Boyanovsky and C.~M.~Ho,
  JHEP {\bf 0707}, 030 (2007);
  Phys.\ Rev.\  D {\bf 76}, 085011 (2007);
  D.~Boyanovsky,
  Phys.\ Rev.\  D {\bf 76}, 103514 (2007);

\bibitem{shi_fuller}
X.~d.~Shi and G.~M.~Fuller,
Phys.\ Rev.\ Lett.\  {\bf 82}, 2832 (1999).

\bibitem{nuMSM}
  T.~Asaka, S.~Blanchet and M.~Shaposhnikov,
  Phys.\ Lett.\ B {\bf 631}, 151 (2005);
  T.~Asaka, M.~Laine and M.~Shaposhnikov,
  JHEP {\bf 0701}, 091 (2007);
  M.~Shaposhnikov,
  Nucl.\ Phys.\  B {\bf 763}, 49 (2007);
  D.~Gorbunov and M.~Shaposhnikov,
  JHEP {\bf 0710}, 015 (2007);

\bibitem{shaposhnikov_tkachev}
  M.~Shaposhnikov and I.~Tkachev,
  Phys.\ Lett.\ B {\bf 639}, 414 (2006).

\bibitem{Kusenko:2006rh}
  A.~Kusenko,
  Phys.\ Rev.\ Lett.\  {\bf 97}, 241301 (2006).

\bibitem{Kadota:2007mv}
  K.~Kadota,
  arXiv:0711.1570 [hep-ph].

\bibitem{pulsars}
  A.~Kusenko and G.~Segr\`e,
Phys.\ Lett.\ B {\bf 396}, 197 (1997);
A.~Kusenko and G.~Segre,
  Phys.\ Rev.\ D {\bf 59}, 061302 (1999).
M.~Barkovich, J.~C.~D'Olivo and R.~Montemayor,
Phys.\ Rev.\ D {\bf 70}, 043005 (2004);
%
G.~M.~Fuller, A.~Kusenko, I.~Mocioiu, and S.~Pascoli,
Phys.\ Rev.\ D {\bf 68}, 103002 (2003);
%
A.~Kusenko,
  Int.\ J.\ Mod.\ Phys.\ D~{\bf 13}, 2065 (2004);
  A.~Kusenko, B.~P.~Mandal and A.~Mukherjee,
  arXiv:0801.4734 [astro-ph].


\bibitem{supernova_misc}
  L.~C.~Loveridge,
  Phys.\ Rev.\  D {\bf 69}, 024008 (2004);
C.~L.~Fryer, A.~Kusenko,
  {\it Astrophys.\ J.\ Suppl.\ } {\bf 163}, 335 (2006);
  J.~Hidaka and G.~M.~Fuller,
  Phys.\ Rev.\  D {\bf 74}, 125015 (2006);
  J.~Hidaka and G.~M.~Fuller,
  Phys.\ Rev.\  D {\bf 76}, 083516 (2007).


\bibitem{reion}
  P.~L.~Biermann and A.~Kusenko,
 Phys.\ Rev.\ Lett.\  {\bf 96}, 091301 (2006);
  M.~Mapelli, A.~Ferrara and E.~Pierpaoli,
  Mon.\ Not.\ Roy.\ Astron.\ Soc.\  {\bf 369}, 1719 (2006);
  J.~Stasielak, P.~L.~Biermann and A.~Kusenko,
  Astrophys.\ J.\  {\bf 654}, 290 (2007);
  E.~Ripamonti, M.~Mapelli and A.~Ferrara,
  Mon.\ Not.\ Roy.\ Astron.\ Soc.\  {\bf 375}, 1399 (2007);
J. Stasielak, P. L. Biermann, A. Kusenko,
arXiv:astro-ph/0701585;
  arXiv:0710.5431 [astro-ph].


\bibitem{Biermann:2007ap}
F. Munyaneza, P.L. Biermann, P. L., Astron and Astrophys., \textbf{436}, 805
(2005);
  P.~L.~Biermann and F.~Munyaneza,
  arXiv:astro-ph/0702173.

\bibitem{x-rays}
K.~Abazajian, G.~M.~Fuller and W.~H.~Tucker,
Astrophys.\ J.\  {\bf 562}, 593 (2001);
  A.~Boyarsky, A.~Neronov, O.~Ruchayskiy and M.~Shaposhnikov,
Mon.\ Not.\ Roy.\ Astron.\ Soc.\  {\bf 370}, 213 (2006);
  A.~Boyarsky, A.~Neronov, O.~Ruchayskiy and M.~Shaposhnikov,
  JETP Lett.\  {\bf 83}, 133 (2006);
 A.~Boyarsky, A.~Neronov, O.~Ruchayskiy, M.~Shaposhnikov and I.~Tkachev,
 Phys.\ Rev.\ Lett.\  {\bf 97}, 261302 (2006);
S.~Riemer-Sorensen, S.~H.~Hansen and K.~Pedersen,
 Astrophys.\ J.\  {\bf 644}, L33 (2006);
  K.~Abazajian and S.~M.~Koushiappas,
  Phys.\ Rev.\  D {\bf 74}, 023527 (2006);
  C.~R.~Watson, J.~F.~Beacom, H.~Yuksel and T.~P.~Walker,
Phys.\ Rev.\  D {\bf 74}, 033009 (2006);
  K.~N.~Abazajian, M.~Markevitch, S.~M.~Koushiappas and R.~C.~Hickox,
Phys.\ Rev.\  D {\bf 75}, 063511 (2007);
  A.~Boyarsky, J.~Nevalainen and O.~Ruchayskiy,
  Astron.\ Astrophys.\  {\bf 471}, 51 (2007);
  A.~Boyarsky, O.~Ruchayskiy and M.~Markevitch,
  arXiv:astro-ph/0611168;
  S.~Riemer-Sorensen, K.~Pedersen, S.~H.~Hansen and H.~Dahle,
  Phys.\ Rev.\  D {\bf 76}, 043524 (2007);
  A.~Boyarsky, J.~W.~den Herder, A.~Neronov and O.~Ruchayskiy,
Astropart.\ Phys.\  {\bf 28}, 303 (2007);
  H.~Yuksel, J.~F.~Beacom and C.~R.~Watson,
  arXiv:0706.4084 [astro-ph];
  A.~Boyarsky, D.~Iakubovskyi, O.~Ruchayskiy and V.~Savchenko,
  arXiv:0709.2301 [astro-ph];
  A.~Boyarsky, D.~Malyshev, A.~Neronov and O.~Ruchayskiy,
  arXiv:0710.4922 [astro-ph].

 \bibitem{viel}
M.~Viel, et al.,
  Phys.\ Rev.\ D {\bf 71}, 063534 (2005);
    %
  U.~Seljak, A.~Makarov, P.~McDonald and H.~Trac,
  Phys.\ Rev.\ Lett.\  {\bf 97}, 191303 (2006)
  %
  M.~Viel, et al., 
A.~Riotto,
  Phys.\ Rev.\ Lett.\  {\bf 97}, 071301 (2006).
  M.~Viel, G.~D.~Becker, J.~S.~Bolton, M.~G.~Haehnelt, M.~Rauch and
W.~L.~W.~Sargent,
  arXiv:0709.0131 [astro-ph].

\bibitem{silk}
  A.~Palazzo, D.~Cumberbatch, A.~Slosar and J.~Silk,
Phys.\ Rev.\  D {\bf 76}, 103511 (2007).


\bibitem{dSphs}
X.~Hernandez, G.~Gilmore, \textit{MNRAS} 297, 517
(1998); J.
Sommer-Larsen, A. D. Dolgov, {\it Astrophys. J.} {\bf 551}, 608
(2001); F. Governato {\it et~al.}, \textit{Astrophys. J.} {\bf 607}, 688
(2004); M.
Fellhauer {\it et~al.}, {\it Astrophys. J.} {\bf 651}, 167 (2006); B. Allgood
\textit{et
al.}, {\it MNRAS} {\bf 367}, 1781 (2006);
T. Goerdt et al., {\em ibid.}, {\bf 368}, 1073 (2006);
G.~Gilmore et al.,
{Astrophys. J.}, {\bf  663}, 948 (2007);
  L.~E.~Strigari, J.~S.~Bullock, M.~Kaplinghat, J.~Diemand, M.~Kuhlen and
P.~Madau,
  arXiv:0704.1817 [astro-ph];
  J.~D.~Simon and M.~Geha,
Astrophys. J. {\bf 670}, 313 (2007);
  R.~F.~G.~Wyse, G.~Gilmore,
  arXiv:0708.1492 [astro-ph];


\bibitem{cdm-wdm}
  G.~Kauffmann, S.~D.~M.~White and B.~Guiderdoni,
  Mon.\ Not.\ Roy.\ Astron.\ Soc.\  {\bf 264}, 201 (1993);
  A.~A.~Klypin, A.~V.~Kravtsov, O.~Valenzuela and F.~Prada,
  Astrophys.\ J.\  {\bf 522}, 82 (1999);
  B.~Moore, S.~Ghigna, F.~Governato, G.~Lake, T.~Quinn, J.~Stadel and P.~Tozzi,
ApJ {\bf 524}, L19 (1999).
  B.~Willman, F.~Governato, J.~Wadsley and T.~Quinn,
MNRAS, 355, 159 (2004);
  P.~Bode, J.~P.~Ostriker and N.~Turok,
  Astrophys.\ J.\  {\bf 556}, 93 (2001).
  P.~J.~E.~Peebles,
ApJ, {\bf 557}, 495 (2001);
  J.~J.~Dalcanton and C.~J.~Hogan,
  Astrophys.\ J.\  {\bf 561}, 35 (2001);
  A.~R.~Zentner and J.~S.~Bullock,
  Phys.\ Rev.\ D {\bf 66}, 043003 (2002);
  J.~D.~Simon, A.~D.~Bolatto, A.~Leroy and L.~Blitz,
  Astrophys.\ J.\  {\bf 596}, 957 (2003);
  F.~Governato {\it et al.},
  Astrophys.\ J.\  {\bf 607}, 688 (2004);
    G.~Gentile, P.~Salucci, U.~Klein, D.~Vergani and P.~Kalberla,
  Mon.\ Not.\ Roy.\ Astron.\ Soc.\  {\bf 351}, 903 (2004);
  J.~Kormendy, M.~E.~Cornell, D.~L.~Block, J.~H.~Knapen and E.~L.~Allard,
  Astrophys.\ J.\  {\bf 642}, 765 (2006);
  M.~I.~Wilkinson {\it et al.},
  arXiv:astro-ph/0602186;
  L.~E.~Strigari, J.~S.~Bullock, M.~Kaplinghat, A.~V.~Kravtsov, O.~Y.~Gnedin,
K.~Abazajian and A.~A.~Klypin, Astrophys.\ J.\  {\bf 652}, 306 (2006);
K.~R.~Stewart, J.~S.~Bullock, R.~H.~Wechsler, A.~H.~Maller and A.~R.~Zentner,
  arXiv:0711.5027 [astro-ph];
  D.~Boyanovsky, H.~J.~de Vega and N.~Sanchez,
  arXiv:0710.5180 [astro-ph].

\bibitem{Gao:2007yk}
  L.~Gao and T.~Theuns,
  Science {\bf 317}, 1527 (2007)

\bibitem{boyan_mixed}
  D.~Boyanovsky,
  arXiv:0711.0470 [astro-ph].

\bibitem{Chikashige:1980ht}
  Y.~Chikashige, G.~Gelmini, R.~D.~Peccei and M.~Roncadelli,
  Phys.\ Lett.\  B {\bf 94}, 499 (1980).
  Y.~Chikashige, R.~N.~Mohapatra and R.~D.~Peccei,
  Phys.\ Lett.\ B {\bf 98}, 265 (1981).

\bibitem{2right-handed}
P.~H.~Frampton, S.~L.~Glashow and T.~Yanagida,
Phys.\ Lett.\ B {\bf 548}, 119 (2002).

\bibitem{Asaka:2006ek}
  T.~Asaka, A.~Kusenko and M.~Shaposhnikov,
  Phys.\ Lett.\ B {\bf 638}, 401 (2006).

\bibitem{McDonald:1993ex}
  J.~McDonald,
  Phys.\ Rev.\  D {\bf 50}, 3637 (1994).

\bibitem{decays}
K.~Sigurdson and M.~Kamionkowski,
  Phys.\ Rev.\ Lett.\  {\bf 92}, 171302 (2004);
  M.~Kaplinghat,
  Phys.\ Rev.\  D {\bf 72}, 063510 (2005)
  J.~A.~R.~Cembranos, J.~L.~Feng, A.~Rajaraman and F.~Takayama,
  Phys.\ Rev.\ Lett.\  {\bf 95}, 181301 (2005).

\bibitem{sterile_constraints}
  A.~Kusenko, S.~Pascoli and D.~Semikoz,
  JHEP {\bf 0511}, 028 (2005).
  A.~Y.~Smirnov and R.~Zukanovich Funchal,
  Phys.\ Rev.\  D {\bf 74}, 013001 (2006).
  G.~Gelmini, S.~Palomares-Ruiz and S.~Pascoli,
  Phys.\ Rev.\ Lett.\  {\bf 93}, 081302 (2004).

\bibitem{deGouvea:2005er}
  A.~de Gouv\^ea,
  Phys.\ Rev.\ D {\bf 72}, 033005 (2005);
  A.~de Gouv\^ea, J.~Jenkins and N.~Vasudevan,
  Phys.\ Rev.\  D {\bf 75}, 013003 (2007).

\bibitem{Candelas:1987rx}
  P.~Candelas and S.~Kalara,
  Nucl.\ Phys.\ B {\bf 298}, 357 (1988).
  D.~Gepner,
  Nucl.\ Phys.\ B {\bf 311}, 191 (1988);
  W.~Buchmuller, K.~Hamaguchi, O.~Lebedev, S.~Ramos-Sanchez and M.~Ratz,
  arXiv:hep-ph/0703078.

\bibitem{Eyton-Williams:2005bg}
  O.~J.~Eyton-Williams and S.~F.~King,
  JHEP {\bf 0506}, 040 (2005).

\bibitem{Mirabelli:1999ks}
N.~Arkani-Hamed, S.~Dimopoulos, G.~R.~Dvali and J.~March-Russell,
  Phys.\ Rev.\  D {\bf 65}, 024032 (2002);
  G.~R.~Dvali and A.~Y.~Smirnov,
  Nucl.\ Phys.\  B {\bf 563}, 63 (1999);
  E.~A.~Mirabelli and M.~Schmaltz,
  Phys.\ Rev.\ D {\bf 61}, 113011 (2000).

\bibitem{baryogenesis}
  E.~K.~Akhmedov, V.~A.~Rubakov and A.~Y.~Smirnov,
  Phys.\ Rev.\ Lett.\  {\bf 81}, 1359 (1998);
  T.~Asaka and M.~Shaposhnikov,
  Phys.\ Lett.\ B {\bf 620}, 17 (2005).

\bibitem{Coleman:1973jx}
  S.~R.~Coleman and E.~Weinberg,
  Phys.\ Rev.\  D {\bf 7}, 1888 (1973).


\bibitem{Dolan:1973qd}
  L.~Dolan and R.~Jackiw,
  Phys.\ Rev.\  D {\bf 9}, 3320 (1974).

\bibitem{Carrington:1991hz}
  M.~E.~Carrington,
  Phys.\ Rev.\  D {\bf 45}, 2933 (1992).

\bibitem{Petraki:2008ef}
  K.~Petraki,
  arXiv:0801.3470 [hep-ph].

\bibitem{Ahriche:2007jp}
  K.~Enqvist, K.~Kainulainen and I.~Vilja,
  Nucl.\ Phys.\ B {\bf 403}, 749 (1993);
  I.~Vilja,
  Phys.\ Lett.\ B {\bf 324}, 197 (1994);
  S.~W.~Ham, Y.~S.~Jeong and S.~K.~Oh,
  J.\ Phys.\ G {\bf 31}, 857 (2005)
  A.~Ahriche,
  Phys.\ Rev.\  D {\bf 75}, 083522 (2007)

\bibitem{McDonald:1993ey_bgen}
  J.~McDonald,
  Phys.\ Lett.\  B {\bf 323}, 339 (1994).

\bibitem{Profumo:2007wc}
 S.~Profumo, M.~J.~Ramsey-Musolf and G.~Shaughnessy,
  JHEP {\bf 0708}, 010 (2007).

\bibitem{Coleman:1977py}
  S.~R.~Coleman,
  Phys.\ Rev.\  D {\bf 15}, 2929 (1977)
  [Erratum-ibid.\  D {\bf 16}, 1248 (1977)].

\bibitem{Linde:1981zj}
  A.~D.~Linde,
  Nucl.\ Phys.\  B {\bf 216}, 421 (1983)
  [Erratum-ibid.\  B {\bf 223}, 544 (1983)].

\bibitem{Kusenko:1995bw}
  A.~Kusenko,
  Phys.\ Lett.\  B {\bf 358}, 47 (1995)
  A.~Kusenko, K.~M.~Lee and E.~J.~Weinberg,
  Phys.\ Rev.\  D {\bf 55}, 4903 (1997).

\bibitem{Weinberg:1987vp}
  E.~J.~Weinberg and A.~q.~Wu,
  Phys.\ Rev.\  D {\bf 36}, 2474 (1987).

\bibitem{Sarid:1998sn}
  U.~Sarid,
  Phys.\ Rev.\  D {\bf 58}, 085017 (1998)
  [arXiv:hep-ph/9804308].



\bibitem{singlet_higgs_LHC}
  T.~Binoth and J.~J.~van der Bij,
  Z.\ Phys.\ C {\bf 75}, 17 (1997);
  A.~Datta et al., 
  Z.\ Phys.\ C {\bf 72}, 449 (1996).
%
  H.~Davoudiasl, T.~Han and H.~E.~Logan,
  Phys.\ Rev.\ D {\bf 71}, 115007 (2005);
  M.~J.~Strassler and K.~M.~Zurek,
  arXiv:hep-ph/0605193;
D.~O'Connell, M.~J.~Ramsey-Musolf and M.~B.~Wise,
  Phys.\ Rev.\  D {\bf 75}, 037701 (2007);
  V.~Barger, P.~Langacker and G.~Shaughnessy,
  Phys.\ Rev.\  D {\bf 75}, 055013 (2007);
  V.~Barger, P.~Langacker, M.~McCaskey, M.~J.~Ramsey-Musolf and G.~Shaughnessy,
  arXiv:0706.4311 [hep-ph].

\bibitem{krs}
V.~A.~Kuzmin, V.~A.~Rubakov and M.~E.~Shaposhnikov,
Phys.\ Lett.\ B {\bf 155}, 36 (1985).


\end{thebibliography}
\end{document}